\documentclass[twocolumn,prl,showpacs,floatfix,superscriptaddress]{revtex4}%
\usepackage{graphicx}
\usepackage{dcolumn}
\usepackage{bm}
\usepackage{amssymb}
\usepackage{amsmath}
\usepackage{amsfonts}
\usepackage{epstopdf}
\usepackage{xcolor}%
\providecommand{\U}[1]{\protect\rule{.1in}{.1in}}
\newcommand{\atio}{A$_2$(Ir$_{1-x}$Ti$_x$)O$_3$}
\newcommand{\natio}{Na$_2$(Ir$_{1-x}$Ti$_x$)O$_3$}
\newcommand{\litio}{Li$_2$(Ir$_{1-x}$Ti$_x$)O$_3$}
\newcommand{\liiro}{Li$_2$IrO$_3$}
\newcommand{\nairo}{Na$_2$IrO$_3$}

\begin{document}
\title{Effect of nonmagnetic dilution in honeycomb lattice iridates Na$_2$IrO$_3$ and Li$_2$IrO$_3$}
\author{S. Manni}
\affiliation{I. Physikalisches Institut, Georg-August-Universit\"at G\"ottingen, 37077 G\"ottingen, Germany}
\author{Y. Tokiwa}\thanks{present address: Research Center for Low Temperature and Materials Science, Kyoto University, Kyoto 606-8501, Japan}
\affiliation{I. Physikalisches Institut, Georg-August-Universit\"at G\"ottingen, 37077 G\"ottingen, Germany}
\date{\today}
\author{P. Gegenwart}
\affiliation{I. Physikalisches Institut, Georg-August-Universit\"at G\"ottingen, 37077 G\"ottingen, Germany}
\affiliation{Experimentalphysik VI, Center for Electronic Correlations and Magnetism, Augsburg University, 86159 Augsburg, Germany}
\date{\today}

\begin{abstract}
We have synthesized single crystals of Na$_2$(Ir$_{1-x}$Ti$_x$)O$_3$ and polycrystals of Li$_2$(Ir$_{1-x}$Ti$_x$)O$_3$ and studied the effect of magnetic depletion on the magnetic properties by measurements of the magnetic susceptibility, specific heat and magnetocaloric effect at temperatures down to 0.1~K. In both systems, the non-magnetic substitution rapidly changes the magnetically ordered ground state into a spin glass, indicating strong frustration. While for the Li system the Weiss temperature $\Theta_{\rm W}$ remains unchanged up to $x=0.55$, a strong decrease $|\Theta_{\rm W}|$ is found for the Na system. This suggests that only for the former system magnetic exchange beyond nearest neighbors is dominating. This is also corroborated by the observation of a smeared quantum phase transition in Li$_2$(Ir$_{1-x}$Ti$_x$)O$_3$ near $x=0.5$, i.e. much beyond the site percolation threshold of the honeycomb lattice.
\end{abstract}

\pacs{75.40.Cx, 75.10.Jm, 75.40.Gb, 75.50.Lk}
\maketitle

\hspace{5.2in}

Iridates have attracted considerable interest in last few years due to their potential to host novel electronic and magnetic phases mediated by the combination of strong spin-orbit (SO) coupling and electronic correlations~\cite{Pesin, Moon, Kim08, Kim09, Okamoto}. Layered honeycomb lattice iridates $A_2$IrO$_3$ ($A=$ Na,Li) are intensively investigated because they have been proposed as candidate materials for the realization of the highly frustrated Kitaev interaction~\cite{Jackeli} as well as correlated topological insulator phases~\cite{Kim, Shitade}.
	
Both Na$_2$IrO$_3$ and Li$_2$IrO$_3$ are electrically insulating with fluctuating $S_{\rm eff}=1/2$ moments above an antiferromagnetic (AF) ordering around 15~K~\cite{Singh10, Singh12}. Their electronic structure is discussed either within $J_{\rm eff}=1/2$ SO Mott insulator~\cite{RIXS} or quasi-molecular orbital (QMO) scenarios~\cite{Mazin12, Foyevtsova2013}, where the upper half-filled $J_{\rm eff}=1/2$ or QMO doublet, respectively, causes magnetism. At present the correct effective Hamiltonians for the description of magnetic exchange in the two systems are not settled. Na$_2$IrO$_3$ displays an AF Weiss temperature of $-120$~K~\cite{Singh10} and zigzag ground state~\cite{Choi12}. Within the next-neighbor Heisenberg-Kitaev (HK) model this would require ferromagnetic (FM) Heisenberg and AF Kitaev couplings~\cite{Jackeli13}, which, however, seems incompatible with ab-initio DFT calculations~\cite{Foyevtsova2013}. Significant further neighbor exchange in a $J_1$-$J_2$-$J_3$ Heisenberg model has been concluded from the analysis of the measured magnon dispersion in Na$_2$IrO$_3$~\cite{Choi12}. On the other hand, it has been pointed out recently, that trigonal distortions present in the system lead to an anisotropic contribution to the next neighbor exchange, which together with a FM Kitaev interaction can reproduce the experimental results~\cite{Imada}.

Isostructural honeycomb Li$_2$IrO$_3$ displays a significantly smaller AF Weiss temperature ($-30$~K) compared to Na$_2$IrO$_3$~\cite{Singh12}. Recent neutron scattering has detected a magnetic Bragg peak within the first Brillouin zone, indicating incommensurate spiral ordering~\cite{Choi14}. Due to the much reduced atomic size of Li, its substitution for Na in (Na$_{1-x}$Li$_x$)$_2$IrO$_3$ revealed that up to $x=0.25$ preferentially only the Na-sites in the honeycomb plane are occupied by Li and further doping results in chemical phase segregation~\cite{Manni}. Magnetic properties of 
Na$_2$IrO$_3$ and Li$_2$IrO$_3$ thus differ significantly~\cite{Manni,Cao}. Due to the smaller Ir-Ir distances in the honeycomb planes in \liiro, one may expect enhanced further neighbor exchange in this system.

Introduction of random vacancies to frustrated magnets induces spin-glass behavior. For striped phases of the HK model it has been shown that the vacancies locally select specific stripe orientations~\cite{Trousselet}. It has recently been proposed, that systematic depletion of the Ir spins by a nonmagnetic ion could provide important new insights on magnetic exchange in these materials. Andrade and Vojta have shown by classical Monte-Carlo simulations that the spin-glass freezing temperatures for depleted next neighbor HK and $J_1$-$J_2$-$J_3$ Heisenberg magnets behave significantly different when the doping concentrations exceed the site percolation threshold $x_p=0.303$~\cite{Vojta}. While in the former case the freezing temperature rapidly drops to zero, spin-glass ordering has a tail and can largely extend into the regime $x>x_p$ for substantial further neighbor magnetic exchange.

We have studied Na$_2$(Ir$_{1-x}$Ti$_x$)O$_3$ and Li$_2$(Ir$_{1-x}$Ti$_x$)O$_3$ where magnetic Ir$^{4+}$ is randomly substituted by nonmagnetic Ti$^{4+}$. In contrast to the Na-system, for the Li-system the AF Weiss temperature remains almost unchanged and spin-glass freezing is found up to $x=0.55$, highlighting the importance of further neighbor exchange in the latter system.

We have chosen nonmagnetic Ti as substituent because Ti$^{4+}$ and Ir$^{4+}$ have a very similar ionic radius. In compounds where Ir and Ti occupy different sites this causes a severe problem due to site exchange~\cite{Dey12}, while in our case it assures a good statistical mixing of Ir and Ti in the diluted systems.
 \natio~ single crystals were grown using a similar method as for \nairo, by pre-reacting Na$_2$CO$_3$, Ir metal powder and TiO$_2$ powder at 750$^{\circ}$C to 900$^{\circ}$C. The subsequent crystal growth was done with 10\% extra IrO$_2$ in between 1030-1050$^{\circ}$C. Unfortunately, this method only worked for compositions $x\leq 0.3$. At larger $x$ only a solid melt of Na$_2$TiO$_3$ was obtained and no \natio~crystals formed. Na$_2$TiO$_3$ has a very low melting point of 180$^{\circ}$C which causes this problem for $x>0.3$. Since chemistry and crystal structure of Na$_2$TiO$_3$ differs from \nairo, attempts to synthesize single-phase \natio~polycrystals for $x>0.3$ have failed.

For \litio~we have prepared well ordered single phase polycrystals up to x=0.55 by solid state reaction. At higher doping \litio~polycrystals become disordered probably due to a site exchange between Li and Ti. For polycrystal synthesis Li$_2$CO$_3$, Ir metal powder and TiO$_2$ were mixed and reacted in the open furnace at 700-1000$^{\circ}$C in 100$^{\circ}$C steps after repetitive grinding and pelletizing after each step. Phase purity and structural ordering were verified from powder x-ray diffraction (XRD). The detailed structural analysis of \litio~by powder XRD (see supplemental information (SI)~\cite{SI}) shows that the changes in the lattice parameters are within 1\% as expected because the ionic radius of Ti$^{4+}$ and Ir$^{4+}$ are similar. For the elemental quantification of the Ir and Ti content several spots on various pieces of each batch have been studied by the energy dispersive x-ray (EDX) method. Throughout this rapid communication $x$ always denotes the actual Ti concentration. Magnetization, ac susceptibility and specific heat measurements were conducted in the Quantum Design MPMS and PPMS. Thermodynamic measurements below 0.4 K were performed in a dilution refrigerator~\cite{Tokiwa}.

\begin{figure}[ht]
\centering
\includegraphics[width=0.9\columnwidth]{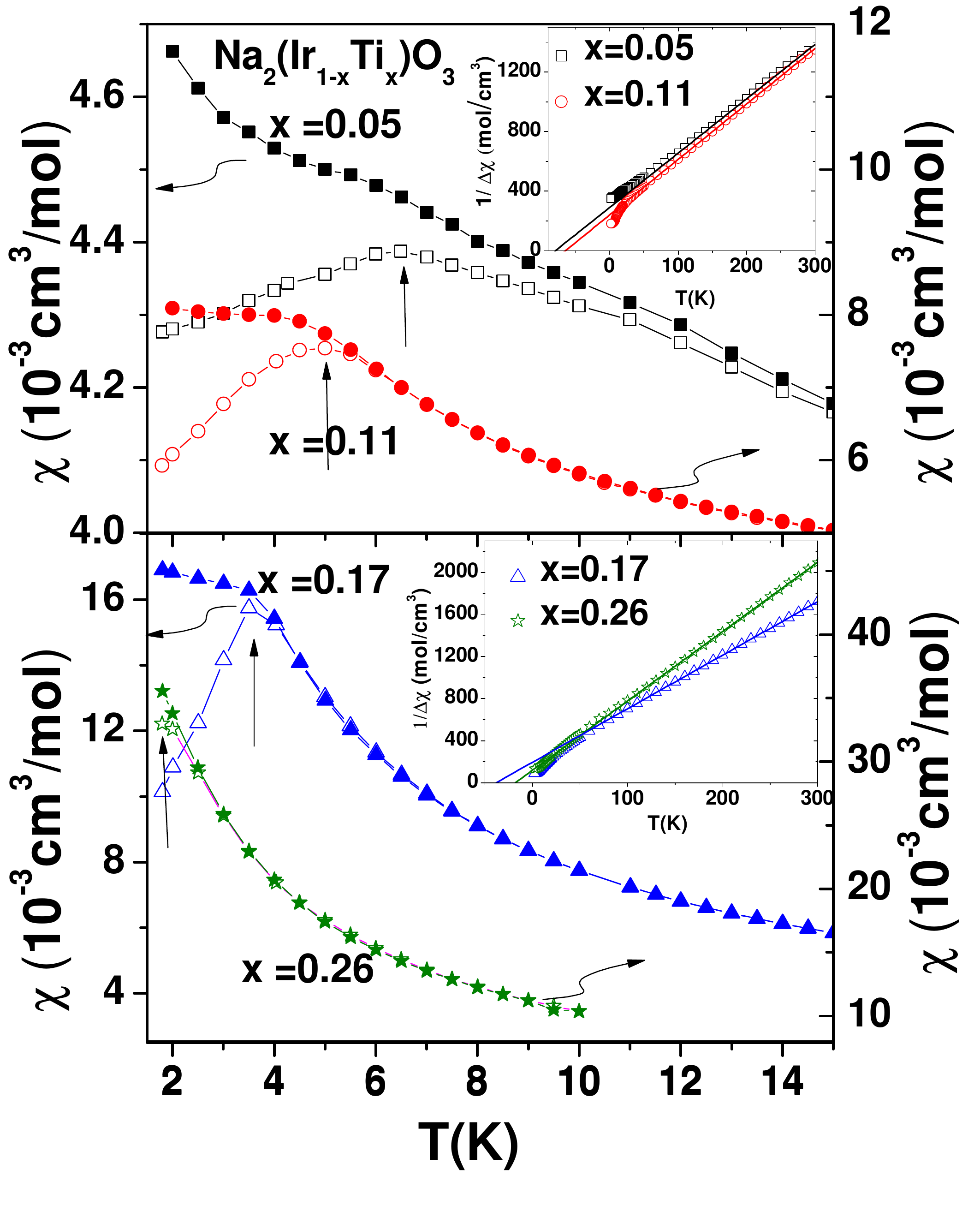}
\caption{Field cooled (FC) and zero field cooled (ZFC) susceptibility vs. temperature as indicated by filled and open symbols, respectively, for \natio~ with $x= 0.05$, $0.11$ (top) and $0.17$, $0.26$ (bottom). Vertical arrows mark T$_g$. Respective insets display $1/\Delta\chi$ (with $\Delta\chi =\chi-\chi_{0}$) vs. $T$. Solid lines indicate Curie-Weiss behavior.} \label{na_mag}
\end{figure}

Magnetization measurements on \natio~single crystals show that for all investigated $x$ the magnetic susceptibility $\chi = M/H$ follows Curie-Weiss (CW) behavior $\chi= \chi_{0} +\frac{C}{T-\theta_{W}}$, see insets of Fig.~\ref{na_mag}. This implies that with increasing degree of dilution by Ti substitution the local moment behavior persists and the decrease of the Curie constant is compatible with the dilution of Ir moments by nonmagnetic Ti (see SI).  Small temperature independent van Vleck contributions ($\chi_{0}$) are of order 10$^{-5}$ cm$^3$/mol. The AF Weiss temperature changes from $-125$~K at $x=0$ to $-18$~K for $x=0.26$ indicating a continuous decrease of the CW scale with magnetic depletion for the Na-system.

Field-cooled (FC) and zero-field cooled (ZFC) measurements at very low field of 5 mT shown in Fig.~\ref{na_mag} display cusps for ZFC and a clear separation between FC and ZFC traces at low $T$, which are characteristic signatures for spin-glass (SG) behavior. The freezing temperature $T_{\rm g}$ has been determined from the maxima in ZFC traces, as indicated by vertical arrows in Fig.~\ref{na_mag}. For the lowest doping level ($x=0.015$) in \natio~long range magnetic ordering is still present below 15~K~\cite{SI}. For higher doping we find a reduction of $T_{\rm g} = 6.8$~K for $x=0.05$ to 2~K for $x=0.26$. The SG behavior is also confirmed by frequency dependent ac susceptibility measurements for $x=0.17$ which show a sharp cusp at $T_{\rm g}$ and a pronounced frequency dependence in the position of that cusp~\cite{SI}. We have also measured the heat capacity ($C$) for this concentration and found a broad hump in $C/T$ above $T_{\rm g}$, which confirms the absence of long-range ordering and indicates SG freezing~\cite{SI}.  
	
\begin{figure}[ht]
\centering
\includegraphics[width=0.9\columnwidth]{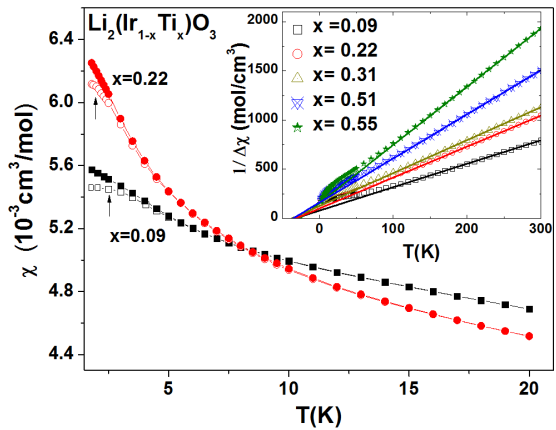}
\caption{FC and ZFC susceptibility (represented by filled and open symbols, respectively) for \litio~with $x= 0.09$ and $0.22$, measured at $H=0.01$~T. Vertical arrows mark $T_{\rm g}$. The inset displays $1/\Delta\chi$ versus $T$ for all investigated $x$. Solid lines illustrate CW behavior.} \label{li_mag}
\end{figure}

Next we discuss the effect of non-magnetic depletion for the Li-system. As shown in Fig.~\ref{li_mag}, \litio~polycrystals display CW behavior between 100 and 300~K (cf. inset). Here $\chi_{0}$ ranges between $-1 \cdot 10^{-5}$ cm$^3$/mol and $-5 \cdot 10^{-5}$ cm$^3$/mol. Remarkably, the observed Weiss temperatures are very similar for all different investigated samples. For $x=0.55$ we observe $-25$~K, which is close to $-33$~K for $x=0$. Hence the CW scale remains almost unchanged for more than 50\% dilution of magnetic moments in the Li-system in stark contrast to its drastic reduction found for the Na-system. At low temperatures, a hysteresis between FC and ZFC susceptibility data is found, similar as for the Na-system. Fig.~\ref{li_mag} shows a separation between the FC and ZFC susceptibility which confirms $T_{\rm g}= 3.5$~K and 2~K for $x=0.09$ and $x=0.22$, respectively (vertical arrows in Fig.~\ref{li_mag} indicate $T_{\rm g}$). The ac susceptibility also shows a strong frequency dependence for these two compositions. Similar SG freezing behavior is also present at higher doping below the temperature limit of our SQUID magnetometer (1.8~K) (see below). 
	
\begin{figure}[pt]
\centering
\includegraphics[width=0.9\columnwidth]{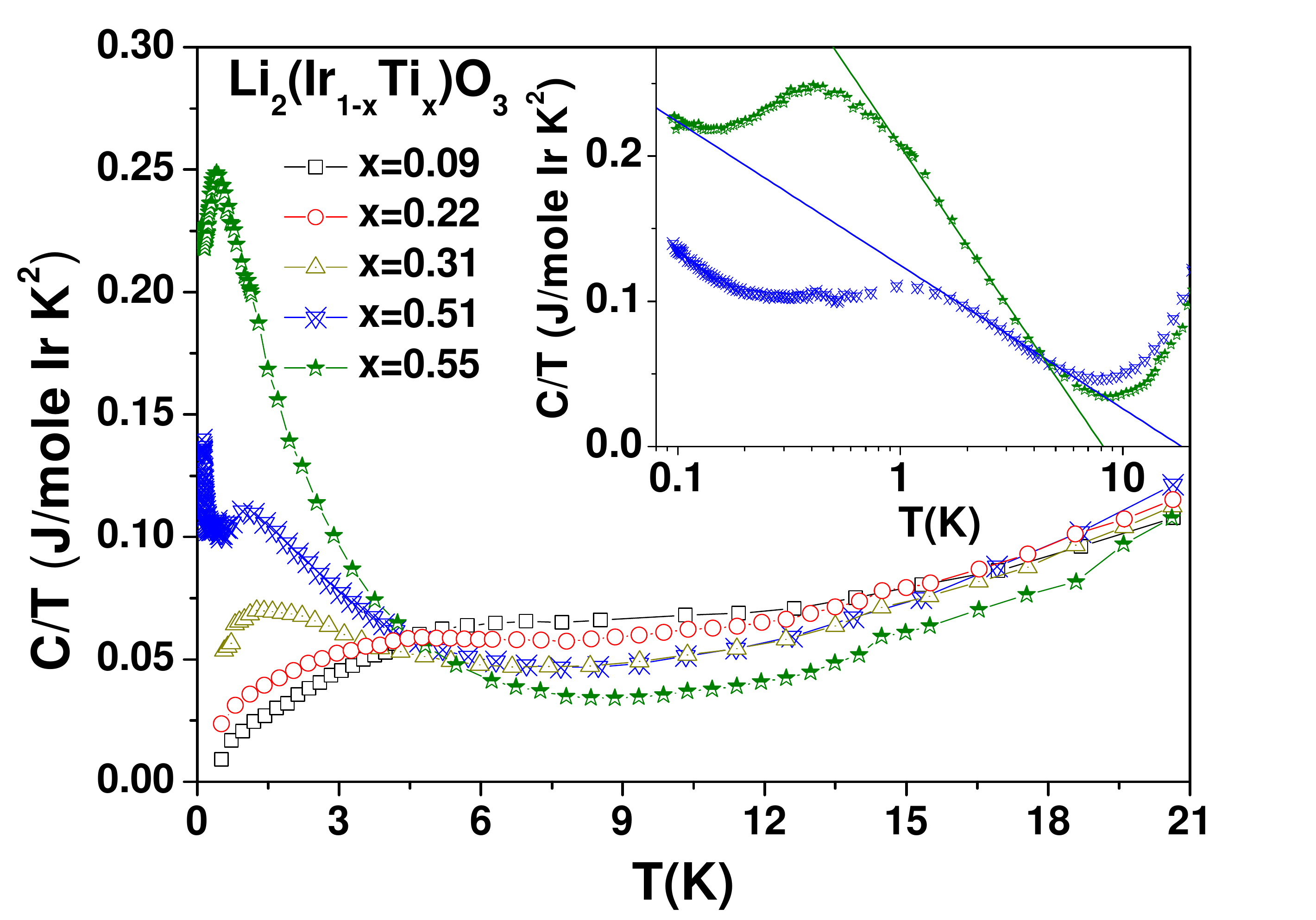}
\caption{Specific heat as $C/T$ vs. $T$ for various \litio~ samples. The inset displays the low-$T$ data for $x=0.51$ and 0.55 vs. $\ln(T)$. Broad maxima indicate $T_{\rm g}$, above which a logarithmic temperature dependence is found (see lines).} \label{liti_C}
\end{figure}

We have measured for all \litio~samples the heat capacity down to 0.4~K and extended the data down to 50~mK for the two highest concentrations, cf. Fig.~\ref{liti_C}. For $x$=0.09 and 0.22 we observe broad maxima in heat capacity divided by temperature  $C/T$ around  $1.4 T_{\rm g}$ which is characteristic for SG transitions (see Fig.~\ref{liti_C}). For $x=0.31$, 0.51 and 0.55 similar broad maxima are found at low temperatures. With increasing $x$ the position of these maxima shift from 1.25~K to 0.41~K for $x=0.31$ to 0.55. The respective $T_{\rm g}$ values are determined by the position of the maximum divided by 1.4. From Fig.~\ref{liti_C} it is unambiguously clear that even beyond 50\% substitution of magnetic Ir sites by non-magnetic Ti in the Li-honeycomb system SG freezing persists and $T_{\rm g}$ continuously shifts to lower temperatures with increasing $x$. Strikingly $C/T$ for $x=0.51$ and $x=0.55$ does not approach 0 at lowest temperatures as expected for insulators but rather saturates (above a low-$T$ nuclear upturn). This implies that a significant amount of magnetic entropy is shifted to low temperatures.

\begin{figure}[pt]
\centering
\includegraphics[width=0.9\columnwidth]{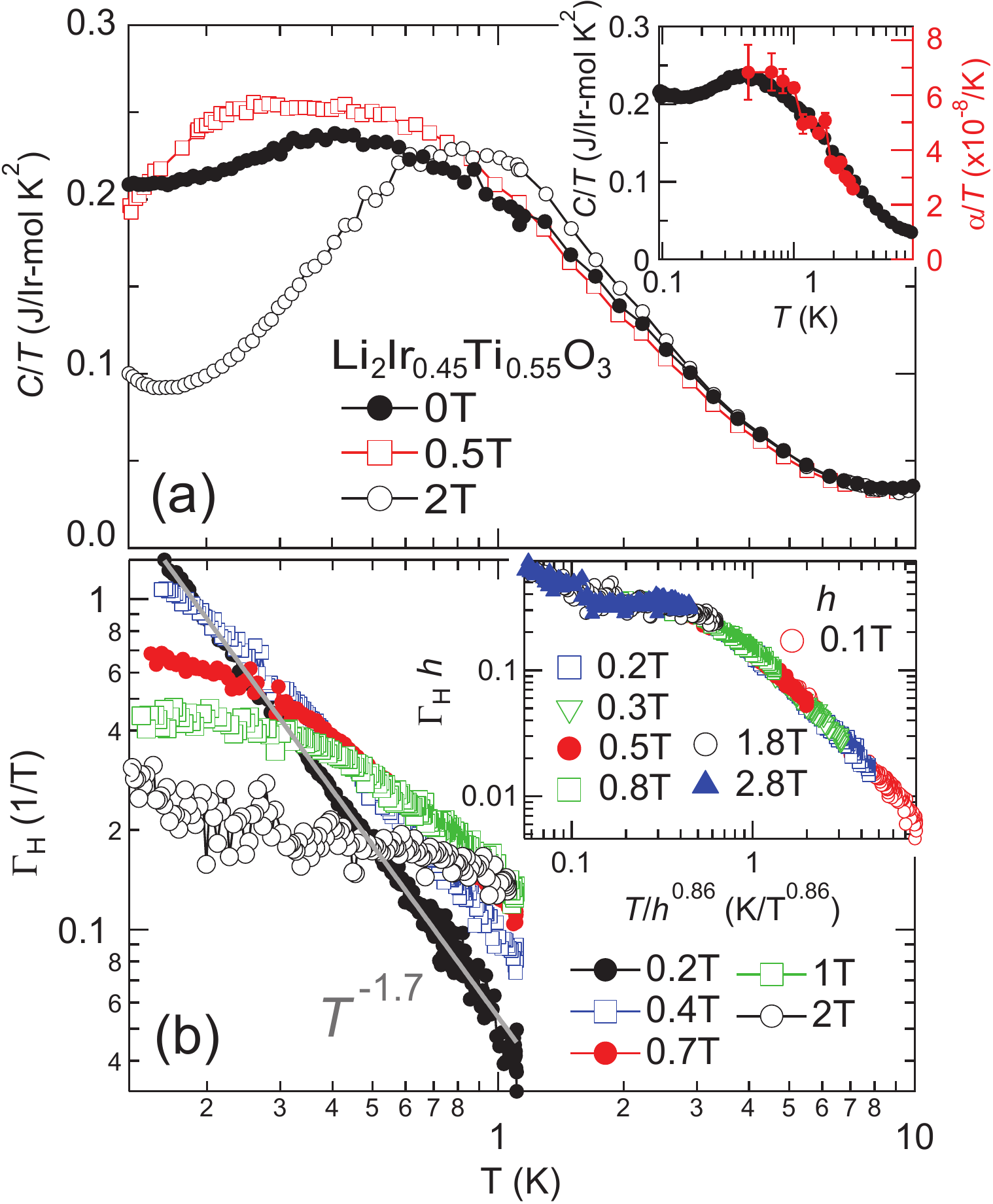}
\caption{Upper panel: Specific heat as $C/T$ vs. $T$ (on log scale) at various fields for \litio, $x=0.55$. The inset displays zero field data together with respective thermal expansion data as $\alpha/T$. Lower panel: Magnetic Gr\"{u}neisen parameter $\Gamma_{\rm H}=T^{-1}(dT/dH)_S$ at different magnetic fields vs. $T$ (on log-log scale). The solid line indicates the $T^{-1.7}$ divergence at 0.2~T. The inset displays scaling behavior,$\Gamma_{\rm H}h$ vs $T/h^\epsilon$ with $\epsilon=0.86$ and $h=(H-0.2$\,T).} \label{liti_qpt}
\end{figure}

As indicated by the straight lines in the inset of Fig.~\ref{liti_C}, a logarithmic increase of $C/T$ is found for $x=0.51$ and $0.55$ upon cooling from about 8~K down to the SG freezing. Such behavior is often found near magnetic instabilities and considered as signature of quantum criticality. We have also observed a strong non-monotonic field dependence of $C/T$ for $x=0.51$\cite{SI} and 0.55 (Fig.~\ref{liti_qpt}).

The adiabatic magnetocaloric effect or magnetic Gr\"{u}neisen parameter $\Gamma_{\rm H}=T^{-1}(dT/dH)_S$ is a sensitive probe of quantum criticality and is expected to diverge as a function of temperature with a power-law function at the critical field $H_c$ for a field-induced quantum critical point (QCP)~\cite{ZhuQCP}. Fig.~\ref{liti_qpt} displays the temperature dependence of $\Gamma_{\rm H}$ at different magnetic fields for \litio, x=0.55. At low field of 0.2~T, a divergence with exponent of $-1.7$ is found over at least one decade in $T$, indicating quantum critical behavior with a low critical field. At 0.4~T and larger fields, $\Gamma_{\rm H}(T)$ saturates upon cooling and the saturation temperature increases with increasing field indicating that fields drive the system away from quantum criticality. The data at various different fields collapse on a single curve when plotted as $\Gamma_{\rm H}h$ vs. $T/h^\epsilon$ (see inset of lower panel of Fig.~\ref{liti_qpt}). Here $h$ denotes the difference in field from the critical field, i.e., $h=H-0.2$~T and the scaling exponent amounts to $\epsilon=0.86$. The critical field of 0.2~T is consistent with the power-law divergence of $\Gamma_{\rm H}(T)$ only observed at 0.2\,T. Furthermore, the non-monotonic field dependence of the low-temperature specific heat is probably due to the small finite $H_c$. A similar divergence and scaling of the magnetic Gr\"uneisen ratio is also found for the $x=0.51$ sample~\cite{SI}. Interestingly, for certain models the possibility of simultaneous percolation and quantum criticality has been investigated theoretically~\cite{Sandvik12, Yu05, Vojta05}.


To further characterize the low-temperature magnetic properties of depleted \litio, we have studied the temperature dependence of the linear thermal expansion coefficient $\alpha(T)=L^{-1}dL/dT$ ($L$: sample length) for $x=0.55$, see upper inset of Fig.~\ref{liti_qpt}. The large values of order $10^{-6}$K$^{-1}$ around 1~K must originate from the magnetic properties (the phonon contribution is several orders of magnitude smaller). Interestingly, $\alpha/T$ perfectly scales with $C/T$ indicating a temperature independent thermal Gr\"uneisen ratio $\Gamma\sim \alpha/C$. This proves the absence of a QCP as function of pressure~\cite{ZhuQCP} and resembles the case of CePd$_{1-x}$Rh$_x$ where $\Gamma(T)$ also does not diverge due to the smeared quantum phase transition (QPT)~\cite{PhilippQGP}. The observed entropy accumulation at low-$T$ which is quenched by a magnetic field but remains unaffected by pressure (or changes in composition) would then arise from weakly coupled magnetic clusters. Our low-temperature experiments on \litio~thus prove that SG formation survives upon substantial magnetic depletion up to $x=0.55$ leading to a smeared QPT.

\begin{figure}[pt]
\centering
\includegraphics[width=0.9\columnwidth]{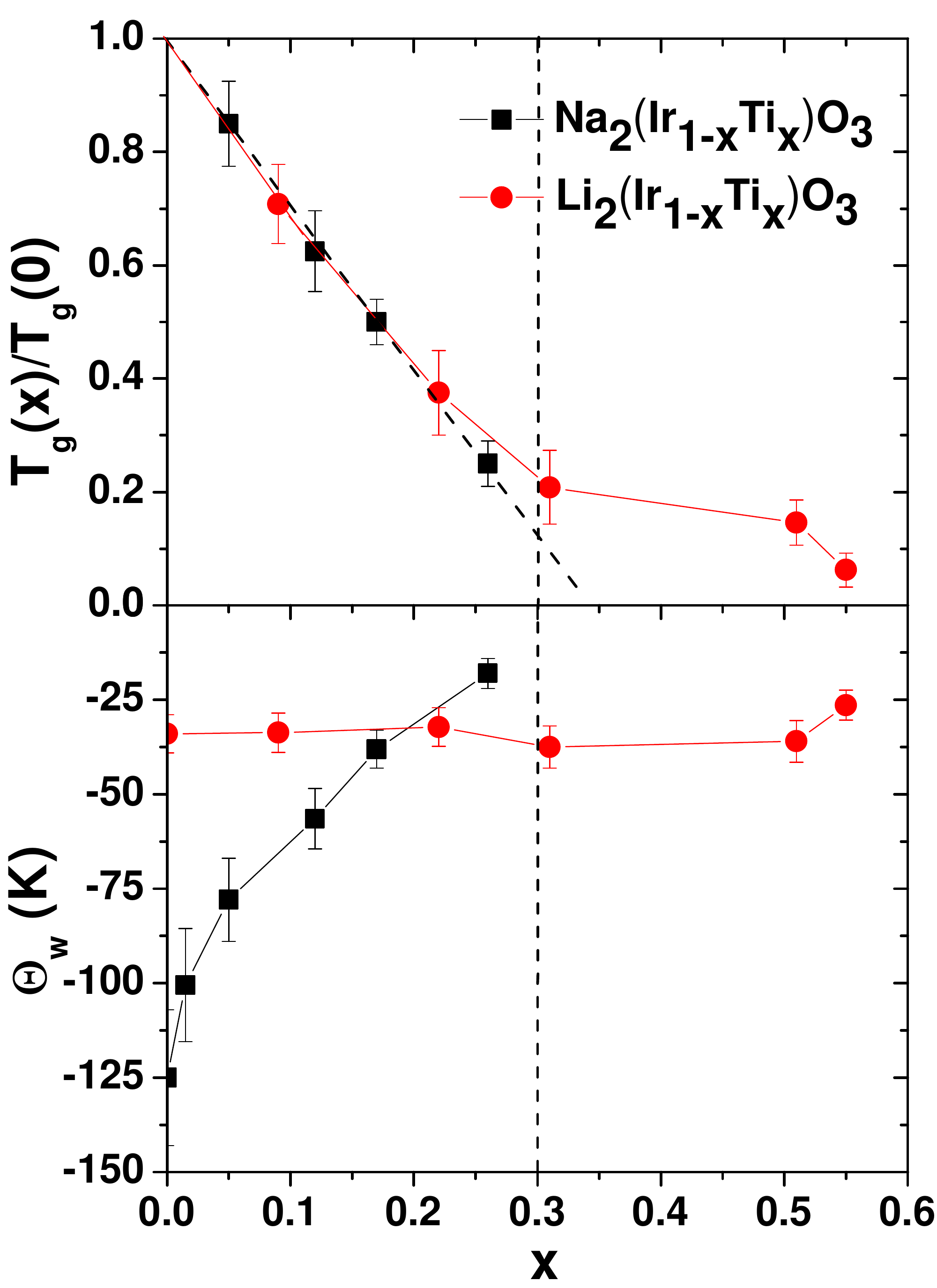}
\caption{Evolution of normalized spin glass ordering temperatures (top) and Curie Weiss temperatures (bottom) for \natio~and~\litio. The vertical dotted line at $x=0.3$ indicates the percolation threshold in the honeycomb lattice. The dashed black line in the upper panel indicates the linear suppression of $T_{\rm g}$ for \natio.} \label{phase}
\end{figure}

The variation of the SG freezing temperatures for the depleted Na- and Li-systems is summarized in the upper panel of Fig.~\ref{phase}. In both systems, already small magnetic depletion induces a SG transition, highlighting the importance of magnetic frustration, and the freezing temperature $T_{\rm g}$ displays a linear suppression at low Ti concentration $x$. However, the evolution of the Weiss temperature shown in the lower panel indicates a substantially different response to magnetic depletion of the two systems. While for the Na-system a drastic reduction of $|\Theta_{\rm W}|$ indicates a suppression of the average magnetic couplings by dilution, $|\Theta_{\rm W}|$ remains unchanged in case of the Li-system. In addition, for the Li-system the signatures of SG formation extend to large $x\sim 0.55$, where signatures of a smeared QPT are observed. Although we could not study \natio~at large $x$, the evolution of the magnetic coupling strength (from $|\Theta_{\rm W}|$) suggests that the QPT for this system is located at significantly lower $x$. Recently classical Monte-Carlo simulations on depleted next neighbor HK and $J_1$-$J_2$-$J_3$ Heisenberg models found that in the former case SG freezing disappears beyond the site percolation threshold $x_p=0.3$ while in the latter case with substantial further neighbor couplings it persists much beyond $x_p$~\cite{Vojta}. Comparison with our data suggests that \nairo~is governed dominantly by the nearest neighbor HK model whereas for the Li-system interactions beyond nearest neighbor are significantly important. Interestingly, $x=0.50$ is the site percolation threshold for a triangular lattice and for $J_2$ exchange only, the honeycomb system corresponds to two decoupled triangular lattices. Thus, the observed smeared QPT must be associated with further neighbor interactions. We also note, that recent theoretical work related to Li$_2$IrO$_3$  found that the low-$Q$ spiral ordering in combination with the AF Weiss temperature $\Theta_W=-30$~K requires a model with second neighbor Kitaev and Heisenberg interactions~\cite{Rachel14}. 

To summarize, we have found differing behaviors in depleted honeycomb Na$_2$IrO$_3$~and Li$_2$IrO$_3$ which suggests the importance of substantial further neighbor magnetic interactions for Li$_2$IrO$_3$. In ~\litio~SG freezing persists to a regime at $x\sim 0.55$ for which indications of a smeared quantum phase transition is observed. Magnetism in this interesting regime could be further investigated by NMR, $\mu$SR or neutron scattering.  
 
We thank Eric C. Andrade, Matthias Vojta and Yogesh Singh for fruitful discussion and collaboration and acknowledge financial support by the Helmholtz Virtual Institute 521 ("New states of matter and their excitations").

\newpage

\section{SUPPLEMENTAL INFORMATION}

\begin{figure}[ptb]
\centering
\includegraphics[width=0.9\columnwidth]{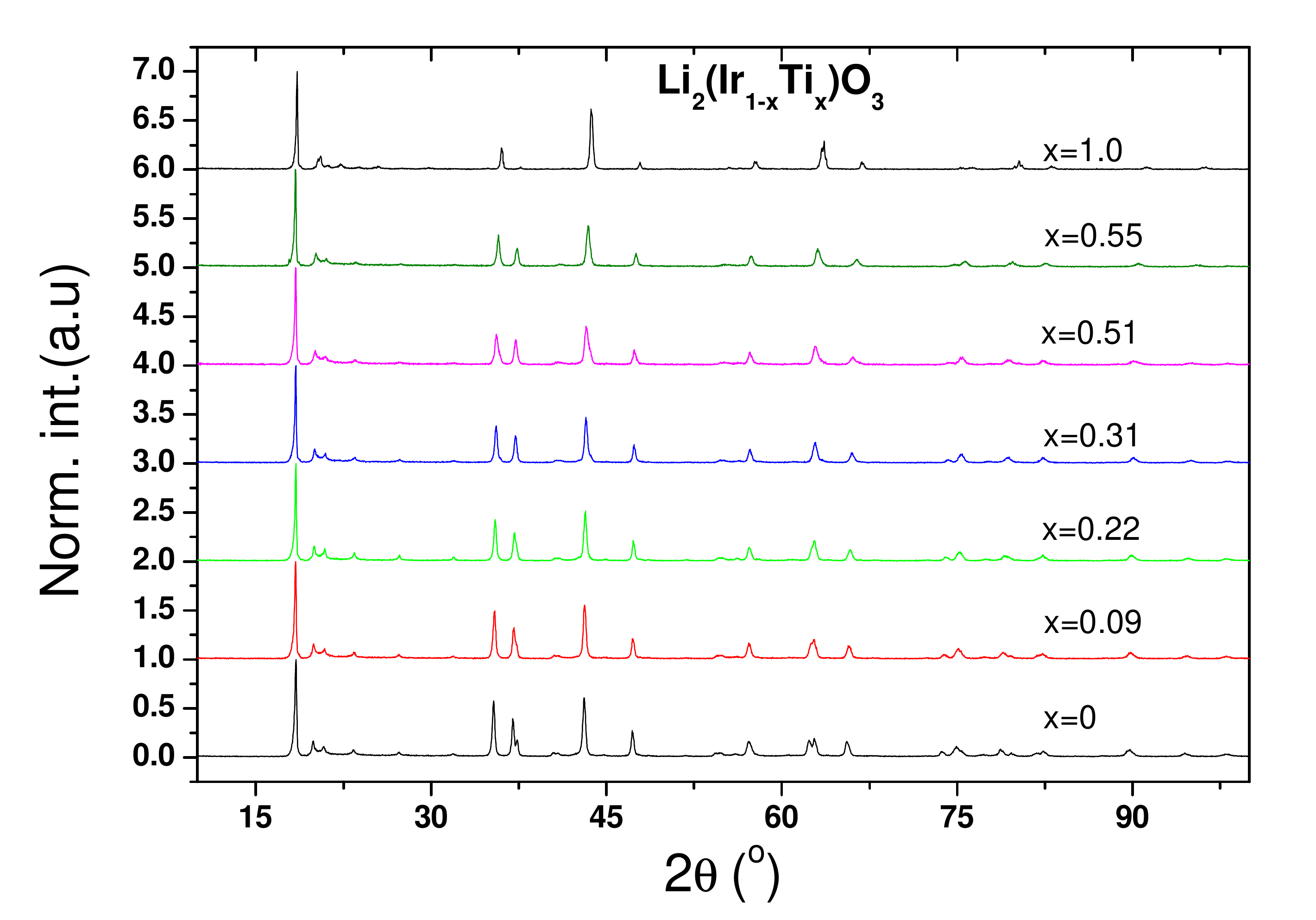} 
\renewcommand\thefigure{S\arabic{figure}}
\caption{Powder XRD patterns of various \litio~polycrystals between $x=0$ and $x=1$. }
\label{liti_xrd}
\end{figure}

\begin{figure}[ptb]
\centering
\includegraphics[width=0.9\columnwidth]{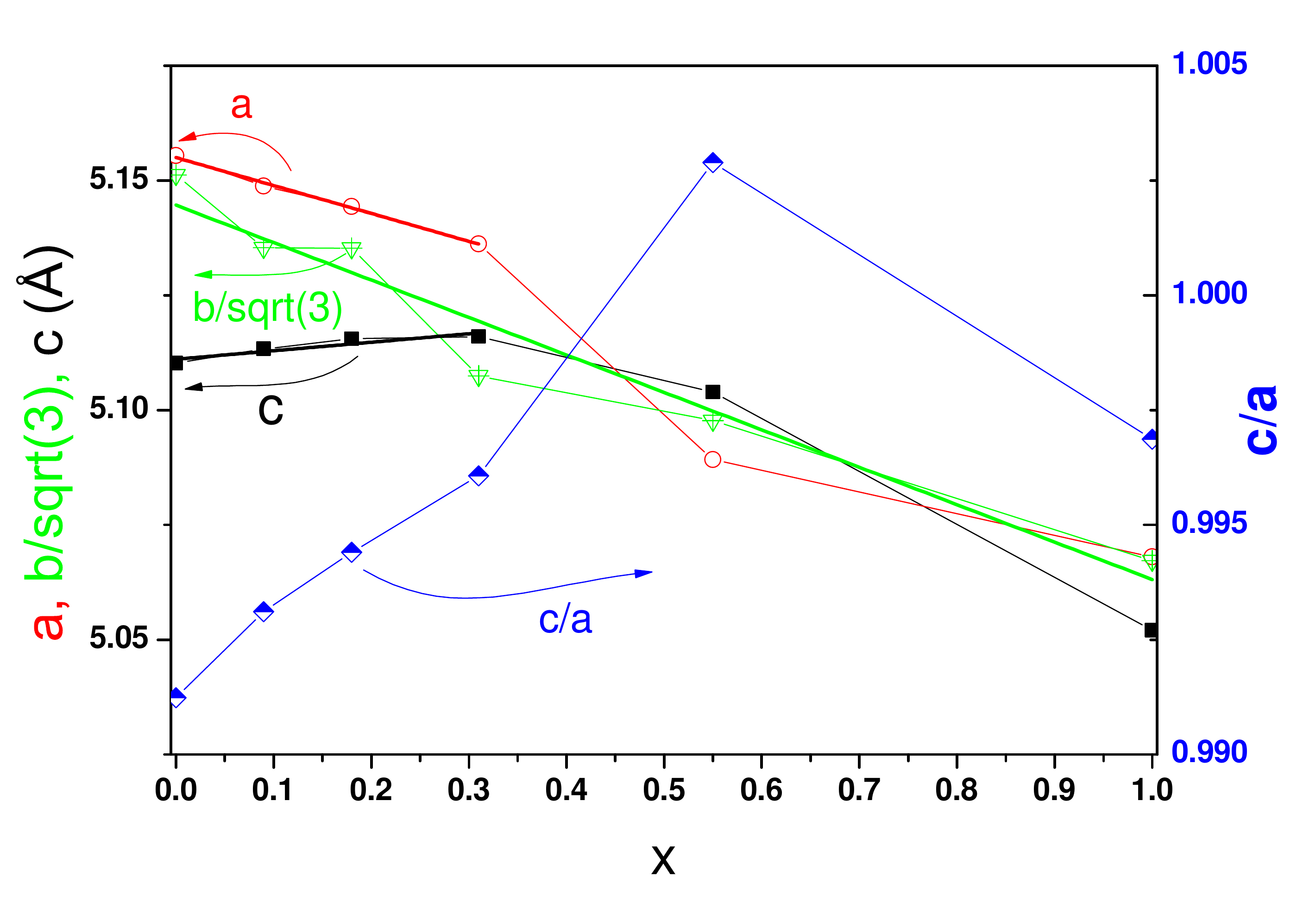}
\renewcommand\thefigure{S\arabic{figure}}
\caption{Variation of lattice parameters of \litio.}
\label{liti_lattice}
\end{figure}

{\it SI. 1. Structural Characterization of \litio:}
For \litio~we did powder XRD measurements for each composition $x$ and fitted the data within the C2/m crystal structure by the Rietveld method. For x$\leq$~0.55 structurally well ordered samples have been obtained as evidences by sharp XRD peaks between $2\theta$ values in the range 19 to 33$^{\circ}$, see Fig.~\ref{liti_xrd}. All XRD peaks for x$\leq$0.55 are matching with \liiro ($x=0$) XRD peaks. However, for x=1, i.e. Li$_2$TiO$_3$, the XRD peaks could not fitted within the C2/m crystal structure, but rather in the C2/c crystal structure.
Lattice parameters of \litio~are plotted in Fig.~\ref{liti_lattice}. They are calculated by structural refinement of the XRD spectra using the Rietveld method. In the structural refinement it was assumed that Ti occupies only the Ir site. The refinement converged and we obtained good fit parameters ($R_p$ and $R_{wp}$) only when we introduced a small site exchange ($f>0$) between the Ir site and the Li site in the honeycomb center. Hence the occupancy at the iridium site is $(1-f-x)$Ir + $x$Ti + $f$Li and at the honeycomb center site it is $(1-2f)$Li + $2f$Ir to balance the stoichiometry. Lattice parameters of \liiro~and Li$_2$TiO$_3$ vary only from 0.5 to 1\% (Li$_2$TiO$_3$ lattice parameters are transformed into C2/m from C2/c for comparison). The $b$ lattice constant changes almost linearly with $x$ from 0 to 1 and $a$ and $c$ vary non-linearly. It is confirmed that by Ti substitution there is no change in crystal structure only a small change in lattice parameters.

{\it SI. 2. Variation in Curie constant in \atio:}
 \begin{figure}[ptb]
\centering
\includegraphics[width=0.9\columnwidth]{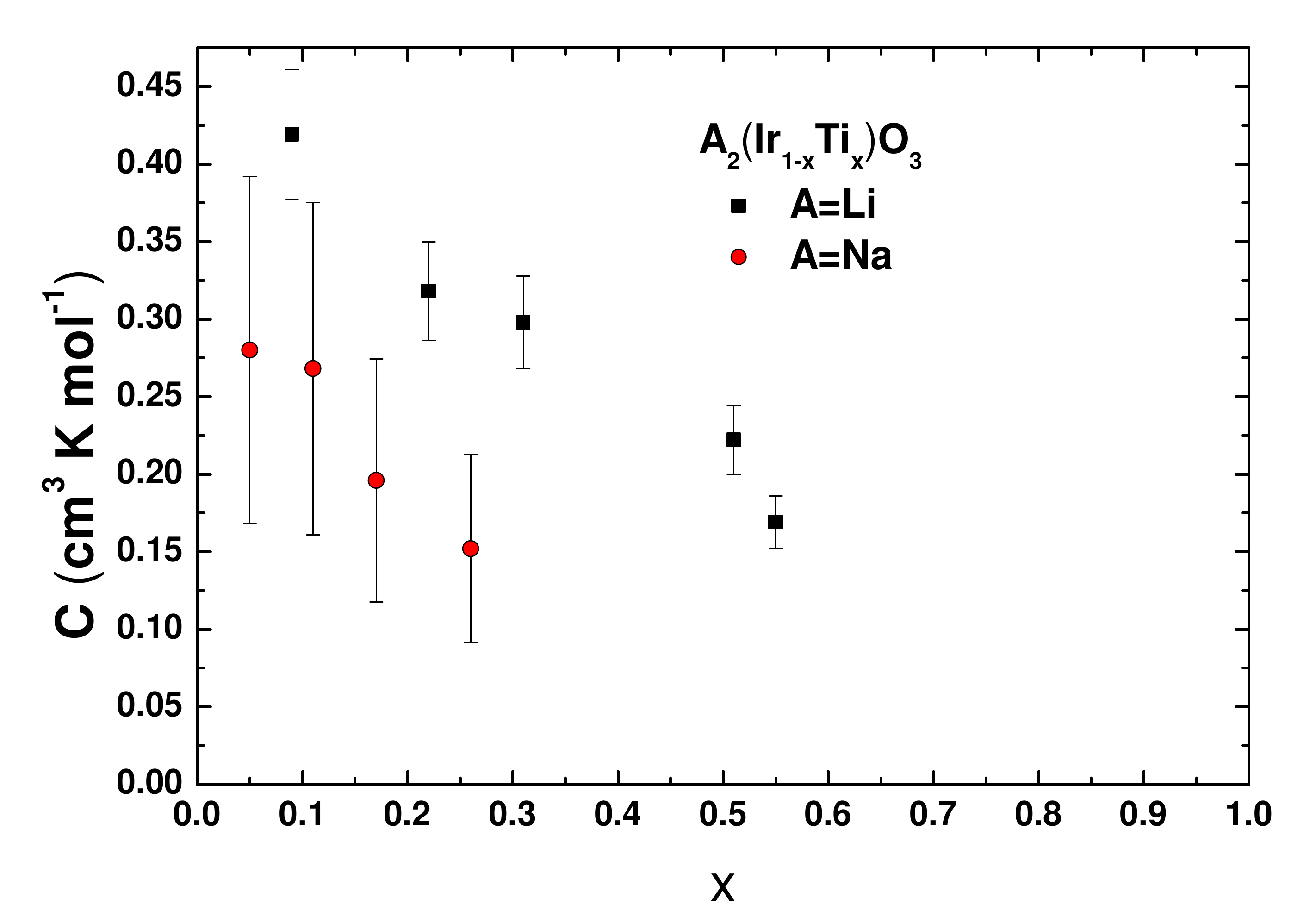}
\renewcommand\thefigure{S\arabic{figure}}
\caption{Variation of Curie constant ($C$) with $x$ in case of \natio~and \litio.}
\label{curie}
\end{figure} 

In the Fig.~\ref{curie} variation of Curie constant ($C$) with $x$ is plotted for \natio~and \litio~which are obtained from the CW fitting of respective $1/\Delta\chi$ versus $T$ data shown in the insets of the Fig.~1~and~2 in the main text. Sharp decrease in $C$ with increasing $x$ confirms dilution of Ir-magnetism for both the Na- and Li-system . Large error bars are used in case of the Na-system to take into account the possible uncertainty in the $C$ value,  due to $g$-factor anisotropy~\cite{Singh10} (magnetization measurement in the Na-system is done on the lump of crystals).

\begin{figure}[ptb]
\centering
\includegraphics[width=0.9\columnwidth]{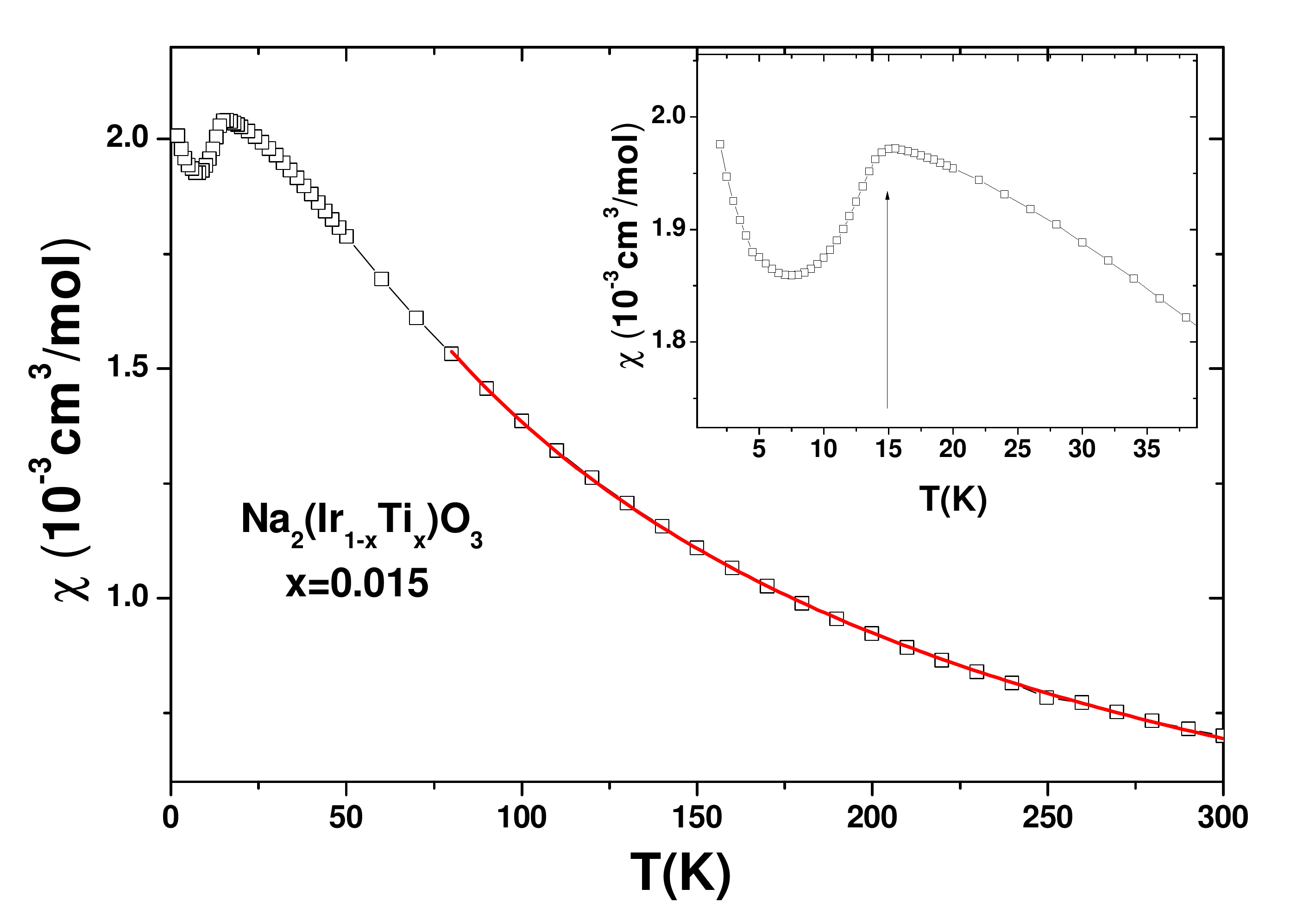}
\renewcommand\thefigure{S\arabic{figure}}
\caption{Magnetic susceptibility ($\chi$) vs. $T$ for \natio single crystal $x=0.015$. The red line indicates Curie-Weiss behavior.  (inset) Zoomed view on low temperature peak, vertical arrow marks $T_N=15$~K.}
\label{nati_ordered}
\end{figure}

{\it SI. 3. Long Range ordering in \natio~,x=0.015:} 
In Fig.~\ref{nati_ordered} we display $\chi$ versus $T$ for the lowest Ti-substituted \natio~single crystal with $x=0.015$. We observe an anomaly with similar shape as for undoped \nairo~\cite{Singh10} at 15~K. We do not observe any hysteresis in field cooled and zero-field cooled data, so the drop in susceptibility signifies long-range AF ordering below 15~K. The red straight line in the figure represents the CW behavior with $\theta_W=-101$~K. 

\begin{figure}[ptb]
\centering
\includegraphics[width=0.9\columnwidth]{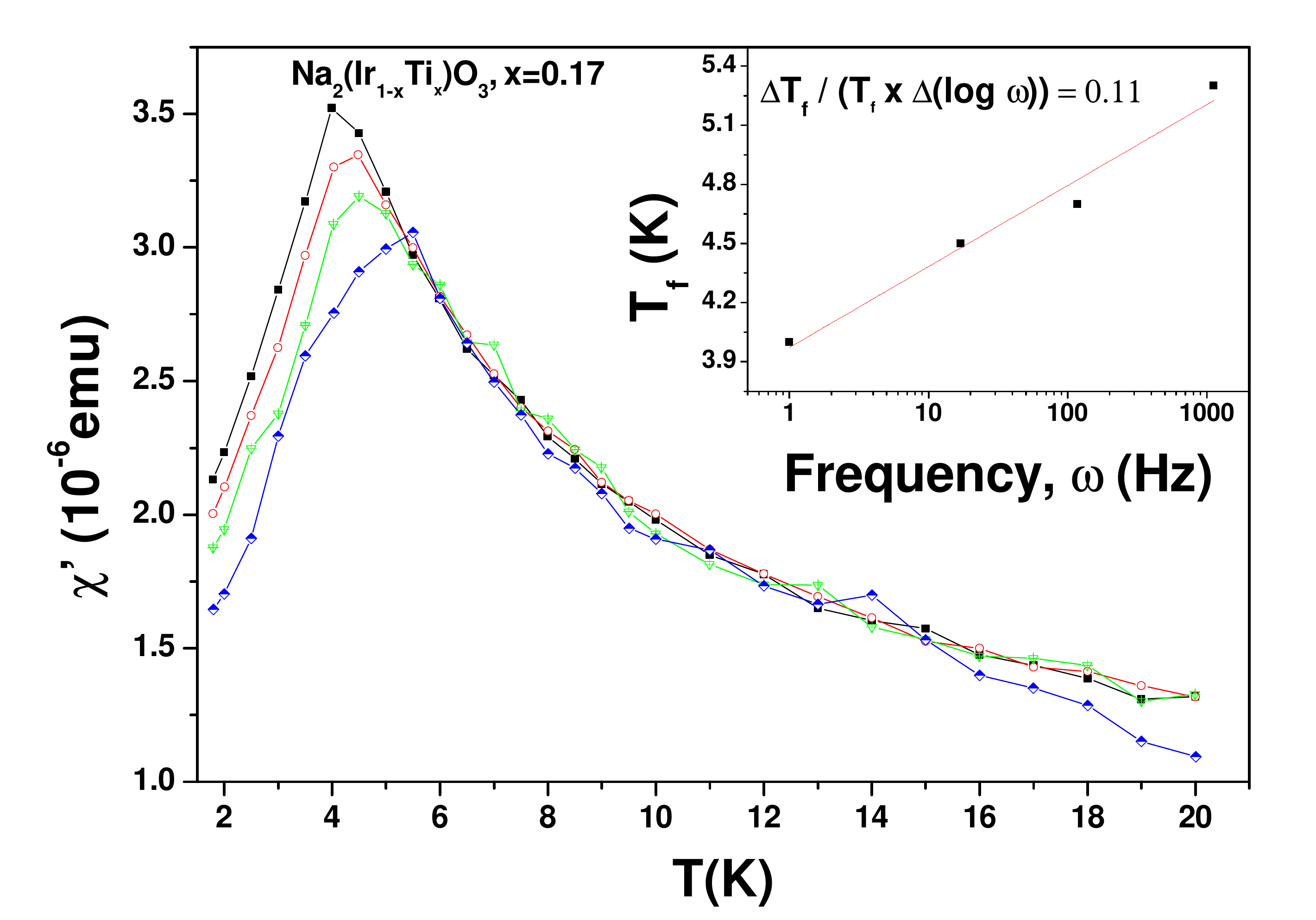}
\renewcommand\thefigure{S\arabic{figure}}
\caption{Temperature dependence of the ac susceptibility ($\chi'$) versus T at various frequencies ($\omega$) near $T_g$. The inset shows frequency dependence (on log scale) of the cusp position $T_f$. Its linear behavior is indicated by the red solid line, see text.}
\label{nati_ac}
\end{figure}

{\it SI. 4. Spin-glass freezing of \natio: } SG feature of \natio was further confirmed from frequency dependent ac susceptibility measurement. Fig.\ref{nati_ac} shows frequency dependent ac susceptibility ($\chi'$) versus T for $x=0.17$ which shows a frequency dependent sharp cusp at $T_g$. At 1 Hz the cusp is very sharp and having peak at $T_g$, with the increasing frequency this cusp shifts towards higher temperature and also slightly broadens with increasing frequency.  To measure this frequency dependence $T_f$ at different frequency ($\omega$) is plotted in logarithmic scale in the inset of Fig.~\ref{nati_ac}. Frequency dependence is measured by $\Delta T_f/(T_g \Delta \log\omega)$ which measures change of $T_g$ per decade of frequency divided by $T_g$. For canonical spin-glasses this value is around 0.02. But as \litio is an insulating system having concentrated local moments, the frequency dependence of $T_g$ is one order of magnitude larger, i.g. around 0.11.

\begin{figure}[ptb]
\centering
\includegraphics[width=0.9\columnwidth]{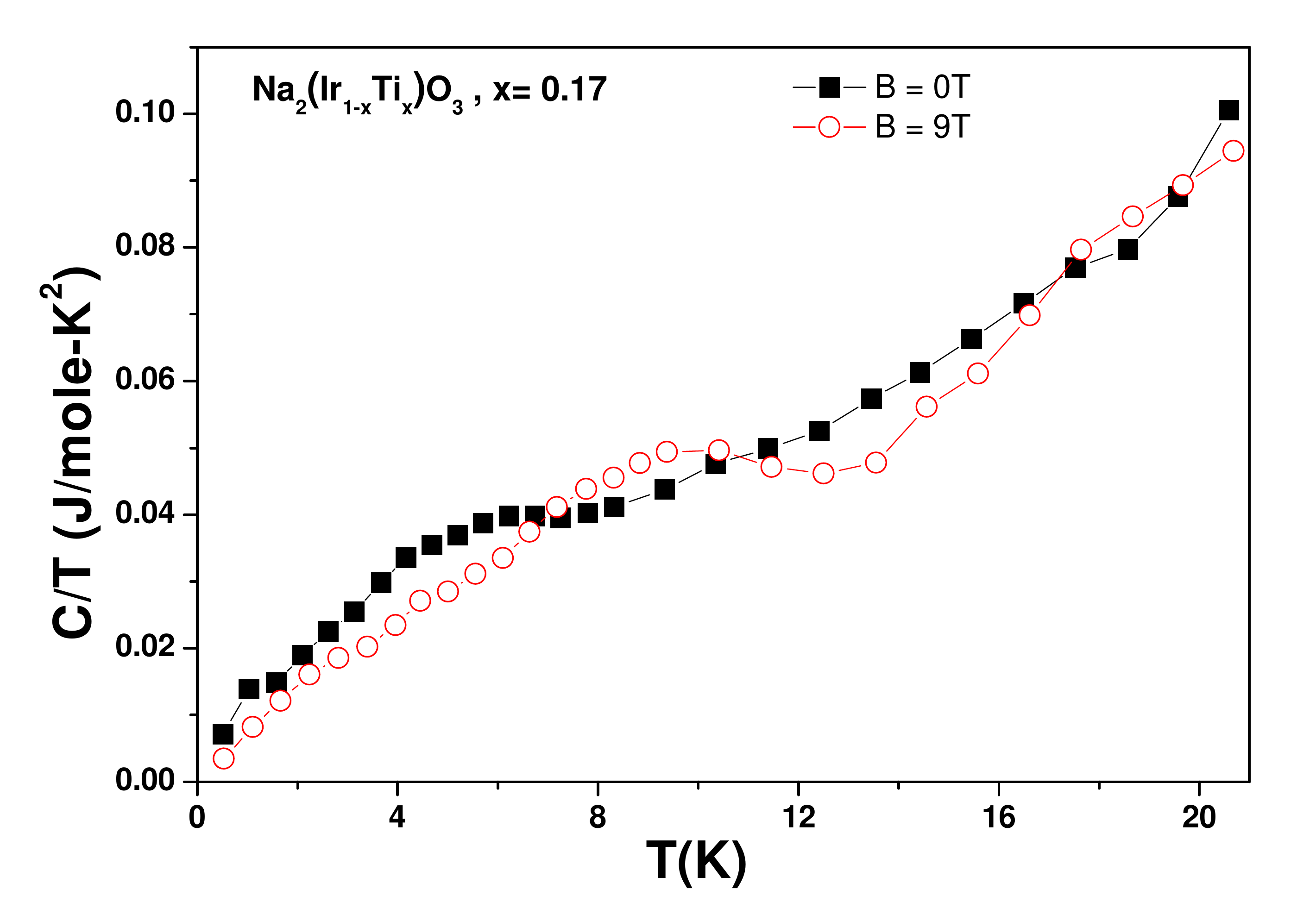}
\renewcommand\thefigure{S\arabic{figure}} 
\caption{Heat capacity divided by temperature ($C/T$) measured at $B=0$ and 9~T for \natio, $x=0.17$.}
\label{nati_Hc}
\end{figure}

To confirm that for $x\geq 0.05$ there is no long range magnetic order the heat capacity has been measured. Crystals were very tiny and hence it was pretty challenging to obtain good data. In fig.~\ref{nati_Hc} the heat capacity divided by temperature ($C/T$) is plotted for \natio~, $x=0.17$ at $B=0$ and 9~T. It shows a broad hump around 6~K which is 1.4 times its $T_g$ determined from ac-susceptibility. Upon applying 9~T magnetic field it broadens and shifts to higher temperatures. Such behavior is characteristic  for spin-glass freezing and it confirms the absence of long-range order.

{\it SI. 5. Adiabatic magnetocaloric effect measurements for \litio, $x=0.51$:}

\begin{figure}[ptb]
\centering
\includegraphics[width=0.8\columnwidth]{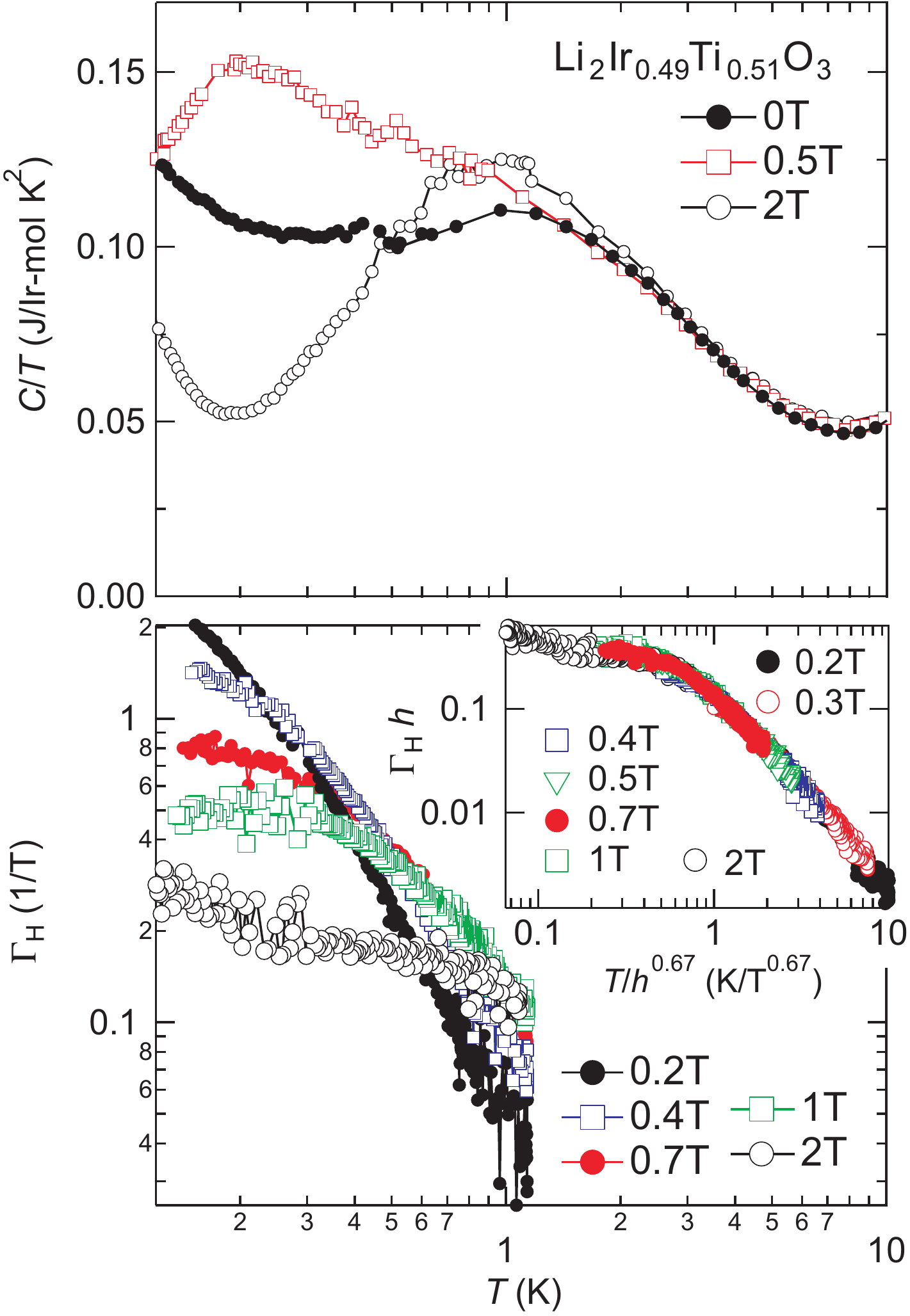}
\renewcommand\thefigure{S\arabic{figure}} 
\caption{(Upper) variation of $C/T$ vs. $T$ for $x= 0.51$,\litio at different $H$. Lower: Magnetic Gr\"{u}neisen parameter $\Gamma_H$ vs. $T$ on log-log scale,  (inset) scaling ,$\Gamma_Hh$ vs $T/h^\epsilon$ plot, with $\epsilon=0.67$ and $h=(H-0.24)$\,T.}
\label{liti_qpt2}
\end{figure}

For \litio, $x=0.51$ the specific heat divided by temperature $C/T$ shows a similar non-monotonic (as for $x=0.55$) but stronger field dependence at low temperature. We observe also same divergence in $\Gamma_H$. Fig.~\ref{liti_qpt2} shows temperature dependence of $\Gamma_H$ at different magnetic fields. $\Gamma_H$ at 0.2\,T diverges as a function of temperature, indicating a presence of QCP very near to this magnetic field. At higher fields, $\Gamma_H$ saturates at low temperatures, suggesting that the system is driven away from QCP at fields above 0.2\,T (lower plot in Fig.~\ref{liti_qpt2}). Similar as for $x=0.55$ , the $x=0.51$ -$\Gamma_H$ data collapse in a common curve, when $\Gamma_Hh$ is plotted against $T/h^\epsilon$ (lower inset Fig.~\ref{liti_qpt2} with $\epsilon=0.67$ and $H_c$=0.24\,T.  Hence field-tuned quantum criticality at  $H_c=0.24$\,T is confirmed in the $x=0.51$ sample.

\end{document}